\documentstyle[preprint,prb,aps]{revtex} 
\begin{document}
\draft

\title{Generalization of the density-matrix method
to a non-orthogonal basis}

\author{R. W. Nunes and David Vanderbilt}

\address{Department of Physics and Astronomy,
Rutgers University, Piscataway, NJ 08855-0849}

\date{\today}
\maketitle

\begin{abstract}
We present a generalization of the Li, Nunes and Vanderbilt
density-matrix method to the case of a non-orthogonal
set of basis functions. A representation of the real-space
density matrix is chosen in such a way that only the overlap
matrix, and not its inverse, appears in the energy functional.
The generalized energy functional is shown to be variational
with respect to the elements of the density matrix,
which typically remains well localized.
\end{abstract}


\narrowtext

Recently, the search for so-called ``order-$N$'' methods
(for which the computational effort scales only
linearly with system size $N$) has led to the development of
a number of new real-space
approaches to the solution of the electronic structure
problem \cite {lnv,daw,mauri,ordej}.
These are based either on the use of a localized,
Wannier-like representation of the occupied subspace
\cite{mauri,ordej}, or on the locality of the real-space
density matrix \cite{lnv,daw}. In the latter case an energy
functional is defined such that the variational degrees of
freedom are the matrix elements of the density matrix in a
real-space-localized set of orthonormal orbitals. However,
in many situations it is more convenient to work
with a non-orthogonal basis (e.g., LCAO calculations using
Gaussian orbitals). For that reason, it becomes desirable to
extend the density-matrix based approaches to those cases.

In this paper we show how the approach proposed by
Li, Nunes and Vanderbilt (LNV) \cite{lnv}  can be extended
to a non-orthogonal basis. This is done by introducing a
new quantity $\bar{\rho}=S^{-1} \rho S^{-1}$ (where $S$ is the
overlap matrix $S_{ij}=\langle\phi_i \mid\phi_j \rangle$, and
$\rho=\langle\phi_i \mid\hat{\rho}\mid\phi_j \rangle$
is the density matrix) as an alternative representation
for the density operator, which is shown to have
similar
localization properties as $\rho$.
Using $\bar{\rho}$ we write a generalized expression for the
total energy in which the inverse overlap matrix $S^{-1}$ does
not appear explicitly; moreover the generalized-density-matrix
(GDM) functional is shown
to be variational with respect to $\bar{\rho}$.

First, we briefly review the LNV approach as applied to
an orthonormal basis \cite{lnv}. For simplicity we consider a
tight-binding description of a system formed by replicating a
unit (super)cell containing $N$ atoms with $M$ basis orbitals
per site. For the moment we assume that the basis orbitals
$\{\phi_i\}$ are orthonormal, i.e.,
$\langle \phi_i\mid \phi_j\rangle=\delta_{ij}$.
For the eigenstates of the Hamiltonian,
$\hat{H}\mid\psi_n\rangle = \epsilon_n \mid\psi_n\rangle$,
we write
\begin{equation}
\mid\psi_n\rangle=\sum_i c_{ni}\mid \phi_i\rangle\;.
\label{egvct}
\end{equation}
The density matrix is defined as
\begin{equation}
\rho_{ij}=\sum_n c_{ni}^* \; c_{nj}^{\vphantom {*}}\;,
\label{dmdef}
\end{equation}
where $i$ and $j$ run over all tight-binding basis orbitals
in the system and $n$ runs over the occupied eigenstates of $H$.
Recall that $\rho$ is a projection onto the subspace of occupied
states, and therefore it obeys the idempotency requirement
$\rho^2=\rho$.

As discussed in Ref. \cite{lnv}, both the standard
$\bf k$-space diagonalization of $H$ and the minimization of
the grand potential
$\Omega = {\rm tr} \left [ \rho \left (H-\mu\right )\right ]$
($\mu$ is a chemical potential used to eliminate the particle
number constraint $N_e = {\rm tr} \left [ \rho\right ]$) with
respect to $\rho$ amount to an ${\cal{O}} (N^3)$ operation.
In the latter case this a result of the idempotency constraint.

In order to achieve an ${\cal{O}} (N)$ solution to the problem,
LNV use the following strategy. First, they take advantage of
the fact that the density matrix is local in real space
\cite{decay,blount,kohn_loc} (in the sense that
$\rho_{ij}\rightarrow0$
as $R_{ij}\rightarrow \infty$, where $R_{ij}$ is the distance
between basis orbitals $\phi_i$ and $\phi_j$), and introduce
a trial density matrix $X$ which is set to zero for
$R_{ij}>R_c$ ($R_c$ is chosen large enough to get a good
approximation to the true density matrix). Second,
no idempotency constraint is explicitly imposed; rather,
they make use of the purification transformation proposed by
McWeeney \cite{mcweeny}:
\begin{equation}
\rho = 3X^2-2X^3 \;.
\label{rho_til}
\end{equation}
This transformation is such that a matrix which is nearly
idempotent ($\lambda_X=1+\delta$ or $\delta$,
$|\delta|<<1$, where $\lambda_X$ is an eigenvalue of
$X$) transforms into a matrix which is more nearly
idempotent
[$\lambda_{\rho}=1-{\cal{O}} (\delta^2)$ or
$+{\cal{O}} (\delta^2)$].
Then, $\rho$ is treated as a physical density
matrix (i.e., ${\rm tr}[\rho A]$ gives the physical
expectation value of operator $A$) and $X$ as a
trial density matrix whose elements constitute the
variational degrees of freedom to be determined by
minimization of the grand potential
\begin{equation}
\Omega = {\rm tr}\left [\rho\left (H-\mu\right )
\right ] = {\rm tr}\left [\left (3X^2-2X^3\right )
\left (H-\mu\right )\right]
\label{omega}
\end{equation}
with respect to $X$. As shown in Ref. \cite{lnv},
$\Omega$ in Eq.\ (\ref{omega}) has a variational local
minimum (i.e., $\Omega\geq\Omega_{\rm exact}$) at which
$\rho$ is idempotent to second order. Since the number
of degrees of freedom per atom is fixed by $R_c$ and no
diagonalization or orthonormalization step is performed,
the above procedure amounts to an ${\cal{O}} (N)$ solution
to the problem.

We now extend the LNV energy functional to a non-orthogonal
basis $\{\phi_i\}$, with the overlap matrix given by
$S_{ij}=\langle \phi_i\mid \phi_j\rangle$. In what follows we
use $X_{ij}=\langle \phi_i \mid \hat{\rho} \mid \phi_j \rangle$
for the trial density matrix in the non-orthogonal basis, to be
consistent with the notation introduced above.
The eigenstates of $\hat{H}$ are given by Eq.\ (\ref{egvct})
and the coefficients $\{c_{ni}\}$ are determined by solving
the secular equation
\begin{equation}
\sum_j\left (H_{ij}-\epsilon_n S_{ij}\right)\;c_{nj}=0\;,
\label{secular}
\end{equation}
where $H_{ij}=\langle\phi_i \mid\hat{H}\mid\phi_j \rangle$.

Let ${\bf C}$ be the matrix defined by ${\bf C}_{in}=c_{ni}$
(i.e., ${\bf C}$ has the eigenvectors $\{\psi_n\}$ as its
columns); it then follows that ${\bf C}$ defines
a congruence transformation that diagonalizes $H$, $S$ and $X$
simultaneously:
\begin{eqnarray}
&&{\bf C}^\dagger H{\bf C}=\Lambda\;,\nonumber \\
&&{\bf C}^\dagger S{\bf C}=I\;,\nonumber \\
&&{\bf C}^\dagger X {\bf C}=X_H^{\vphantom {*}}\;,
\label{transf}
\end{eqnarray}
where $I$ is the identity matrix, and
$\Lambda_{mn}=\epsilon_n\delta_{mn}$ and
$\left (X_H^{\vphantom {*}}\right )_{mn}=
\theta(\mu-\epsilon_n)\delta_{mn}$ are, respectively,
the matrices of $\hat{H}$ and $\hat{\rho}$ in the basis
$\{\psi_n\}$ of the eigenvectors of $H$
[$\theta(x)$ is the theta function].
{}From  Eq.\ (\ref{transf}) we have
\begin{equation}
S^{-1}={\bf C}{\bf C}^\dagger\;.
\label{sinv}
\end{equation}

Using Eqs.\ (\ref{transf}) and  (\ref{sinv}), and
$\rho_H^{\vphantom {*}} = 3 X_H^2- 2 X_H^3$
following Eq.\ (\ref{rho_til}),
we can now generalize Eq.\ (\ref{omega}) as follows:
\begin{eqnarray}
\Omega&&={\rm tr}\left [\rho_H^{\vphantom {*}}
\left (\Lambda-\mu\right )\right ] \nonumber \\
&&={\rm tr}\left [\left (3S^{-1}X S^{-1}X S^{-1}
-2S^{-1}X S^{-1}X S^{-1}X S^{-1} \right )
H^\prime\right ]\;,
\label{energ_1}
\end{eqnarray}
where $H^\prime =H-\mu S$.

As a matter of convenience, we would like to eliminate
$S^{-1}$ from the energy expression in favor of $S$. This can
be accomplished by defining the two new quantities
\begin{eqnarray}
\bar{X} &=& S^{-1}X S^{-1}\;, \nonumber \\
{\bar{\rho}} &=& 3\bar{X} S\bar{X}
-2\bar{X} S\bar{X} S\bar{X}\;,
\label{rho_overline}
\end{eqnarray}
as alternative representations \cite{dual} for the trial
and physical density matrices, respectively.
We observe that $\bar{X}$ is a more natural representation
of the density operator, in the
sense that Eq.\ (\ref{dmdef}) still holds, i.e.,
$\bar{X}_{ij}=\sum_n c_{ni}^* c_{nj}^{\vphantom {*}}$,
whereas
$X_{ij}=\sum_n \sum_{kl} S_{ik}^{\vphantom {*}} c_{nk}^*
c_{nl}^{\vphantom {*}} S_{lj}^{\vphantom {*}}$.
Furthermore, the expectation value of any operator is given by
$\langle \hat{A} \rangle = {\rm tr}\;[\bar{X} A]$, where
$A_{ij} = \langle \phi_i \mid \hat{A} \mid \phi_j \rangle $.

In terms of $\bar{X}$
and $\bar{\rho}$ the particle number becomes
\begin{equation}
N_e={\rm tr}\left [ {\bar{\rho}}S \right ]=
{\rm tr}\left [\left (3\bar{X} S\bar{X}
-2\bar{X} S\bar{X} S\bar{X}  \right )
S\right ]\;,
\label{e_numb}
\end{equation}
and the energy functional is written
\begin{equation}
\Omega={\rm tr}\left [ {\bar{\rho}}H^\prime \right ]=
{\rm tr}\left [\left (3\bar{X} S\bar{X}
-2\bar{X} S\bar{X} S\bar{X}  \right )
H^\prime\right ]\;.
\label{energ_2}
\end{equation}
To show that Eq.\ (\ref{energ_2}) is variational with
respect to $\bar{X}$, we note the following. At the solution,
the density matrix must obey the idempotency constraint
$X_H^2=X_H^{\vphantom {*}}$, which in the new representation
is expressed as
\begin{equation}
\bar{X} S \bar{X} = \bar{X}\;;
\label{idem}
\end{equation}
furthermore, it must also commute with the Hamiltonian, i.e,
$X_H^{\vphantom {*}}\Lambda = \Lambda X_H^{\vphantom {*}}$.
In terms of $\bar{X}$, we have
\begin{equation}
S \bar{X} H = H \bar{X} S\;.
\label{commut}
\end{equation}
Eqs.\ (\ref{idem}) and (\ref{commut}) can easily be obtained
by applying the transformation generated by $\bf C$ to the
more familiar expressions in the basis $\{\psi\}$, and then
using $X = S\bar{X} S$.
{}From Eqs.\ (\ref{idem}) and (\ref{commut}) it follows
immediately that the variational gradient
\begin{equation}
{{\delta\Omega}\over{\delta\bar{X}}}=3\left (S\bar{X}
H^\prime+H^\prime\bar{X}S\right )-2\left (S\bar{X}
S\bar{X}H^\prime+S\bar{X}H^\prime\bar{X}S
+H^\prime\bar{X}S\bar{X}S\right )\;
\label{gradient}
\end{equation}
vanishes at the solution, thus showing that $\Omega$
is variational with respect to $\bar{X}$.

Eqs.\ (\ref{e_numb}), (\ref{energ_2}), and (\ref{gradient})
constitute the central results of this work.  Note that the
standard LNV scheme is recovered from these equations upon
substituting $S_{ij}=\delta_{ij}$.

Before proceeding further, let us comment on the real-space
localization properties of $S$ and $S^{-1}$. We are
interested in the case where the basis orbitals are
localized in real space, and therefore $S$ is also
localized. It can be shown that $S^{-1}$ is then
exponentially localized \cite{s_decay,ziman}, with a decay
length that depends on the spread of the eigenvalues of $S$.
If $S$ is an ill-conditioned matrix, i.e.,
$\max(\lambda_S)/\min(\lambda_S)\gg 1$ [$\max(\lambda_S)$ and
$\min(\lambda_S)$ are, respectively, the maximum and minimum
eigenvalues of $S$], then $S^{-1}$ has a long decay length.

The advantage of using Eq.\ (\ref{energ_2}) is that it
eliminates the need to compute $S^{-1}$. A possible
concern, in making use of the current approach, may be
that the matrix $\bar{X}$ may decay more slowly with distance
than would the density matrix expressed in terms of orthogonal
basis orbitals. This may happen when $S$ is ill-conditioned,
such that $S^{-1}$ has a longer exponential decay than $X$
\cite{remark}. When this is the case, it becomes necessary
to increase the cutoff radius $R_c$ to obtain the same level
of accuracy. However, as we discuss below, our numerical
evidence suggests that $X$ and $\bar{X}$ will, in general,
have very similar decay as each other, and very similar to
that of $X$ for an orthogonal basis, and therefore the
transformation leading from Eq.\ (\ref{omega}) to
Eq.\ (\ref{energ_2}) typically
preserves the localization properties
of $\Omega$. Note also that the presence of $S$ in
Eq.\ (\ref{energ_2}) [as compared to Eq.\ (\ref{omega})] does
not affect the linear scaling of the method, since $S$ is as
localized as $H$ in a local basis.

Next, we present some numerical tests for a three-dimensional
tight-binding (TB) model for silicon, to illustrate the
localization properties of $X$ and $\bar{X}$. We use the
universal TB model proposed by Harrison in its extension to a
non-orthogonal basis \cite{harr}. The matrix elements of $S$ in
a minimal $sp^3$-basis are given by
$S_{ll^\prime n}=2 k V_{ll^\prime n}
(\epsilon_l+\epsilon_{l^\prime})^{-1}$,
where $l$ and $l^\prime$ run over $s$ and $p$ orbitals and $n$
indicates the type ($\sigma$ or $\pi$) of interaction. The
$V_{ll^\prime n}$'s are the universal TB parameters introduced by
Harrison \cite{harr}, $\epsilon_l$ are atomic on-site energies and
$k$ is an adjustable parameter. For the Hamiltonian matrix we have
$H_{ll^\prime n}=(1+k-S_2^2) V_{ll^\prime n}$,
where $S_2$ is the overlap between two $sp^3$ hybrids
$S_2=(S_{ss\sigma}-2\sqrt{3}S_{sp\sigma}-3S_{pp\sigma})/4$.
Both $H$ and $S$ are restricted to first neighbors only. For
simplicity we set $k=1$ which is very close to the value $1.042$
commonly used for silicon.

In Fig.~\ref{fig:rhos} we show the behavior of $X$ and
$\bar{X}$ as functions of the distance $R_{ij}$, between two
orbitals $\phi_i$ and $\phi_j$. Plotted is the norm
$||X^{ij}||=\left[\sum_{\alpha \beta}
{|X^{ij}_{\alpha \beta}|}^2/N\right ]^{1/2}\;$,
where $\alpha$ and $\beta$ run over $\{s,p_x,p_y,p_z\}$ and $N=4$
is the block dimension. Also shown is the behavior of
$X_{\rm ortho}$ for orthogonal orbitals, which is obtained by
setting $k=0$ in the TB model. It can be seen that within a
distance $R_{ij}=8.00$ \AA, both $X$ and $\bar{X}$ as well as
$X_{\rm ortho}$ decay to $\sim2.0$\% of their values at the origin.

We have also calculated $X$ and $\bar{X}$ for a
basis with an ill-conditioned (almost singular) $S$, by setting
$k=$1.35 (this implies a large overlap between
neighboring orbitals). The results are shown in
Fig.~\ref{fig:ill_s}.
Also shown is the long range behaviour of $S^{-1}$.
Although $S^{-1}$ has a very long decay
length in this case, $X$ and $\bar{X}$ still decay very
similarly as $X_{\rm ortho}$. The point here is to show that,
provided that the basis is sufficiently localized,
$\bar{X}$ can be as localized as $X_{\rm ortho}$ for an
orthonormal basis, even when $S^{-1}$ is ill-conditioned.
In any case, by using Eq.\ (\ref{energ_2}), the convergence
of both quantities, $\hat{\rho}$ and $S^{-1}$, is built into
a single quantity $\bar{X}$, and is controlled by a single
parameter $R_c$.

Because of the fact that $S$ is of the same range as $H$
(i.e., $R_S^{\vphantom {*}}=R_H^{\vphantom {*}}$),
the GDM scheme preserves
the ${\cal{O}} (N)$ scaling of the original method.
Nevertheless, the presence of $S$ in Eq.\ (\ref{energ_2})
implies an increase in the scaling pre-factor.
In the orthonormal case, the time-dominant step involves
the calculation of a product
$X_{jk}(X H)_{ki}$ of two matrices of range $R_c$ and
$R_c+R_H^{\vphantom {*}}$, out to a radius $R_{ij}\leq R_c$.
In the non-orthonormal
case, the corresponding dominant operation involves two matrices
$(SX)_{jk}(SX H)_{ki}$ of range $R_c+R_H^{\vphantom {*}}$
and $R_c+2R_H^{\vphantom {*}}$,
calculated also up to the radius $R_c$ [see
Eq.\ (\ref{gradient})].
In order to estimate the slowdown factor,
we determined the ratio of the number of terms that contribute
in each case, which was found to be 3.6 using the $R_c$ and
$R_H^{\vphantom {*}}$ of Ref. \cite{lnv}.

In summary, we have presented a generalization of the LNV
density-matrix approach to the case of a non-orthogonal basis.
An alternative real-space representation
of the density operator is introduced, which is argued to have
similar
localization properties as the conventional
density matrix, as suggested by the numerical evidence presented.
In this generalized energy functional, only the
overlap matrix $S$ appears explicitly (as opposed to $S^{-1}$).
The new functional is shown to retain its variational property
and its linear scaling with system size.

This work was supported by NSF Grant DMR-91-15342.
R. W. Nunes acknowledges the support from the
CNPq - Conselho Nacional de Desenvolvimento Cient\'{\i}fico e
Tecnol\'ogico, Brazil.



\begin{figure}
\caption{
$X$, ${\bar{X}}$ and $X_{\rm ortho}$ as functions
of the distance $R_{ij}$ between two basis orbitals,
for the tight-binding model of the text.
\label{fig:rhos}}
\end{figure}

\begin{figure}
\caption{
Same as before for ill-conditioned $S\;(k=1.35)$;
long range behaviour of $S^{-1}$ is also shown.
\label{fig:ill_s}}
\end{figure}


\begin{references}

\bibitem{lnv} X.-P.~Li, R.~W.~Nunes, and D.~Vanderbilt,
Phys. Rev. B {\bf 47}, 10891 (1993).

\bibitem{daw} M.~Daw, Phys. Rev. B {\bf 47}, 10895 (1993).

\bibitem{mauri} F.~Mauri, G.~Galli, and R.~Car,
Phys. Rev. B {\bf 47}, 9973 (1993); F.~Mauri and G.~Galli,
to be published.

\bibitem{ordej} P.~Ordejon, D.~A.~Drabold, M.~P.~Grumbach,
and R.~Martin, Phys. Rev. B {\bf 48}, 14646 (1993).

\bibitem{decay} The decay is power-law in metals ($R^{-d}$ in
$d$ dimensions), and exponential in insulators; in the latter
case, the decay length is related to that of the Wannier
functions as shown in Refs. \cite{blount,kohn_loc}.

\bibitem{blount} E. I. Blount, in {\sl Solid State Physics,}
edited by F. Seitz and D. Turnbull (Academic, New York, 1962),
Vol. 13, p. 305.

\bibitem{kohn_loc} W.~Kohn, Phys. Rev. {\bf 115}, 809 (1959);
Phys. Rev. B {\bf 7}, 4388 (1973);.

\bibitem{mcweeny} R.~McWeeny, Rev. Mod. Phys. {\bf 32}, 335 (1960).

\bibitem{dual} In terms of the dual orbitals
$\mid {\bar{\phi_i}}\rangle
=\sum_j S_{ji}^{-1}\mid {\phi_j}\rangle$,
we have
${\bar{X}}_{ij}= \langle {\bar{\phi_i}} \mid
{\hat{\rho}} \mid {\bar{\phi_j}} \rangle$.

\bibitem{s_decay} To show that $ S^{-1}$ is exponentially localized,
let the spectrum of $S$ be:
$\min(\lambda_S) \leq \lambda_S \leq \max(\lambda_S)$.
Since $S$ is positive definite, $\min(\lambda_S) > 0$ and
the problem of inverting $S$ can be mapped onto the problem
of finding the Green's function $ G = ( H - \epsilon )^ {-1}$
of a {\it Hamiltonian} matrix $H = S $ at {\it energy}
$\epsilon =0$.  But the Green's function is always
exponentially localized at energies outside the band \cite{ziman},
so $S^{-1}$ is exponentially localized.

\bibitem{ziman} John M. Ziman, {\it Models of disorder}
(Cambridge University Press, 1979, Cambridge, England), Chap. 8.

\bibitem{remark} As an example, for a TB model in which {\it all}
bands are occupied we have $X=S$ and $\bar{X}=S^{-1}$.

\bibitem{harr} W.~Harrison, Phys. Rev. B {\bf 27}, 3592 (1983);
M.~van Schilfgaarde and W.~Harrison, Phys. Rev. B {\bf 33},
2653 (1986);

\end{references}
\end{document}